\documentclass[12pt]{article}
\usepackage{graphicx}
\usepackage{graphics}
\usepackage{amsthm}
\usepackage{amssymb, amsmath}
\title{Series representation of the Riemann zeta function and other results:
Complements to a paper of Crandall} 
\author{Mark W. Coffey\\
Department of Physics\\
Colorado School of Mines\\
Golden, CO  80401\\
(Received $\mbox{~~~~~~~~~~~~~~~~~~~~~~~~~~~~~~~2012}$)}
\date{February 26, 2012}
\begin{document}
\maketitle
\baselineskip=25 pt
\begin{abstract}

We supplement a very recent paper of R. Crandall concerned with the multiprecision
computation of several important special functions and numbers.  We show an alternative
series representation for the Riemann and Hurwitz zeta functions providing analytic
continuation through out the whole complex plane.  Additionally we demonstrate some series
representations for the initial Stieltjes constants appearing in the Laurent expansion 
of the Hurwitz zeta function.  A particular point of elaboration in these developments is
the hypergeometric form and its equivalents for certain derivatives of the incomplete
Gamma function.  Finally, we evaluate certain integrals including
$\int_{\mbox{\tiny{Re}} s=c} {{\zeta(s)} \over s} ds$ and $\int_{\mbox{\tiny{Re}} s=c} {{\eta(s)} \over s} ds$, with $\zeta$ the Riemann zeta function and $\eta$ its alternating
form.

\end{abstract}
 
\medskip
\baselineskip=15pt
\centerline{\bf Key words and phrases}
\medskip 

\noindent

Riemann zeta function, Hurwitz zeta function, Stieltjes constants, Euler polynomials,
Bernoulli polynomials, incomplete Gamma function, confluent hypergeometric function, 
generalized hypergeometric function

\vfill
\centerline{\bf 2010 AMS codes} 
11M06, 11M35, 11Y35, 11Y60
 
\baselineskip=25pt
\pagebreak
\medskip
\centerline{\bf Introduction and statement of results}
\medskip

Very recently R. Crandall has described series-based algorithms for computing multiprecision values of several important functions appearing in analytic number theory and the theory of special functions \cite{crandall2012}.  These functions include the Lerch transcendent 
$\Phi(z,s,a)=\sum_{n=0}^\infty z^n/(n+a)^s$, and many of its special cases, and Epstein zeta functions.  

In the following we let $B_n(x)$ and $E_n(x)$ be the Bernoulli and Euler polynomials,
respectively, $\Gamma$ the Gamma function, and $\Gamma(x,y)$ the incomplete Gamma function (e.g., \cite{nbs,andrews,grad}).  
An example series representation from \cite{crandall2012} (27) is that for the Hurwitz zeta 
function $\zeta(s,a)$,
$$\zeta(s,a)={1 \over {\Gamma(s)}}\sum_{n=0}^\infty {{\Gamma[s,\lambda(n+a)]} \over {(n+a)^s}}
+ {1 \over {\Gamma(s)}}\sum_{m=0}^\infty (-1)^m {{B_m(a)} \over {m!}} {\lambda^{m+s-1} \over
{(m+s-1)}}, \eqno(1.1)$$
with free parameter $\lambda \in [0,2\pi)$.

We complement the presentation of \cite{crandall2012} with further representation of the
Riemann zeta function $\zeta(s)=\zeta(s,1)$ and some results on the Stieltjes constants.
The latter constants $\gamma_n(a)$ (e.g., \cite{coffey2009,coffeyjnt,coffeyseries,coffeysums}) 
appear in the Laurent expansion 
$$\zeta(s,a)={1 \over {s-1}}+ \sum_{n=0}^\infty {{(-1)^n} \over {n!}}\gamma_n(a)
(s-1)^n, \eqno(1.2)$$
where $\gamma_0(a)=-\psi(a)$, the Euler constant $\gamma_0=\gamma=-\psi(1)$, and by convention one takes $\gamma_k = \gamma_k(1)$.  Here $\psi=\Gamma'/\Gamma$ denotes
the digamma function (e.g., \cite{nbs,andrews,grad}).  
The Stieltjes constants may be expressed via the limit formula
$$\gamma_k(a)=\lim_{N \to \infty}\left[\sum_{n=0}^N {{\ln^k(n+a)} \over {n+a}}- {{\ln^{k+1}
(N+a)} \over {k+1}}\right].$$

We let $\eta(s)=(1-2^{1-s})\zeta(s)$ denote the alternating Riemann zeta function, given
for Re $s>0$ by $\eta(s)=\sum_{n=1}^\infty {{(-1)^{n-1}} \over n^s}$.  
The following series representation provides an analytic continuation of the Riemann zeta
function to the whole complex plane.
{\newline \bf Proposition 1}.  Let $|\lambda| <\pi$.  Then
$$\Gamma(s)\eta(s)=\sum_{m=1}^\infty (-1)^{m-1} {{\Gamma(s,m\lambda)} \over m^s}+{1 \over 2}
\sum_{n=0}^\infty {{E_n(0)} \over {n!}} {\lambda^{n+s} \over {(n+s)}}.  \eqno(1.3)$$
This representation holds in all of $\mathbb{C}$.  In particular, it delivers the special
values $\eta(-j)=(-1)^j E_j(0)/2$ for $j \geq 0$.
An alternative expression for the values $E_n(0)$ is
\cite{nbs} (p. 805) $E_n(0)=-2(n+1)^{-1}(2^{n+1}-1)B_{n+1}$ for $n \geq 1$ in terms of 
Bernoulli numbers $B_n=B_n(0)$.  

{\it Remark}.  The values $E_n(0)=-E_n(1)$ for $n \geq 2$ even are zero while $E_0(0)=1$.
The signs of two adjacent values of odd indices are opposite.

Proposition 1 has many implications.  For example, for $\lambda=1$ we have the following,
with Li$_s$ denoting the polylogarithm function.
{\newline \bf Corollary 1}.
$$\ln 2=\ln(1+e)-1+{1 \over 2}\sum_{n=0}^\infty {{E_n(0)} \over {(n+1)!}}, \eqno(1.4a)$$
and
$${1 \over 2}\zeta(2)=\ln(1+e)-1-\mbox{Li}_2\left(-{1 \over e}\right)+{1 \over 2}\sum_{n=0}^\infty {{E_n(0)} \over {n!(n+2)}}.  \eqno(1.4b)$$

We apply the representation (1.1) for a later Proposition.  First, we verify some properties 
of the Hurwitz zeta function from (1.1).  
{\newline \bf Corollary 2}.  (a)
$$\zeta(s,a)=\zeta(s,a+1)+a^{-s}, \eqno(1.5)$$
(b)
$$\partial_a \zeta(s,a)=-s\zeta(s+1,a),  \eqno(1.6)$$
(c) for integers $q \geq 2$, 
$$\sum_{r=1}^{q-1} \zeta\left(s,{r \over q}\right)=(q^s-1)\zeta(s),  \eqno(1.7)$$
and (d) for Re $s<1$,
$$\int_0^1 \zeta(s,a)da=0.  \eqno(1.8)$$

By induction, (1.6) gives $\partial_a^j \zeta(s,a)=(-1)^j (s)_j\zeta(s+j,a)$, where 
$(z)_n=\Gamma(z+n)/\Gamma(z)$ is the Pochhammer symbol.

We let $_pF_q$ denote the generalized hypergeometric function (e.g., \cite{nbs,andrews,grad}).
Derivatives of the incomplete Gamma function with respect to the first argument may be
expressed in hypergeometric form.  An example is given in the following.
{\newline \bf Proposition 2}.
$$\Gamma^{(1)}(s,x) \equiv {\partial \over {\partial s}}\Gamma(s,x)=\int_x^\infty t^{s-1}\ln t
~ e^{-t}dt$$
$$={x^s \over s^2} ~_2F_2(s,s;s+1,s+1;-x)+\Gamma(s)[-\ln x+\psi(s)]+\ln x \Gamma(s,x).  \eqno(1.9)$$

Later we provide separate discussion for such derivatives.

\noindent
{\bf Proposition 3}.  Let Re $a>0$.  Then (a)
$$-\gamma-\psi(a)=\gamma_0(a)-\gamma=e^{-a}\Phi\left({1 \over e},1,a\right)+\sum_{m=1}^\infty
{{(-1)^m} \over {m!}}{{B_m(a)} \over m}, \eqno(1.10)$$
(b)
$${\gamma^2 \over 2}+\gamma \psi(a)+{{\zeta(2)} \over 2}-\gamma_1(a)=\sum_{n=0}^\infty
{{\Gamma(0,n+a)} \over {n+a}}-\sum_{m=1}^\infty {{(-1)^m} \over {m!}} {{B_m(a)} \over m^2},
\eqno(1.11)$$
and (c)
$$-{\gamma^3 \over 6}-\gamma {{\zeta(2)} \over 2}-\left({\gamma^2 \over 2}+{{\zeta(2)} \over 2}
\right)\psi(a)+\gamma \gamma_1(a)-{{\zeta(3)} \over 3}+{{\gamma_2(a)} \over 2}$$
$$=\sum_{n=0}^\infty \left\{- ~_3F_3(1,1,1;2,2,2;-n-a)+{1 \over {(a+n)}}\left[\gamma \ln(n+a)
+{\gamma^2 \over 2}+{{\zeta(2)} \over 2}+{1 \over 2}\ln^2(n+a)\right]\right\}$$
$$+\sum_{m=1}^\infty {{(-1)^m} \over {m!}} {{B_m(a)} \over m^3}.  \eqno(1.12)$$

{\bf Corollary 3}.  For $\lambda \in [0,2\pi)$,
$$\ln \Gamma(x)=\gamma (1-x)-\ln \lambda (x-1)+
+\sum_{n=0}^\infty\left\{\Gamma[0,\lambda(n+x)]-\Gamma[0,\lambda(n+1)]\right\}$$
$$+\sum_{m=2}^\infty {{(-1)^m} \over {(m-1)m!}}[B_m(x)-B_m]\lambda^{m-1}.$$

Alternative forms of the Bernoulli polynomial sums of Proposition 3 and of another sum in
(1.11) are given in the following.
{\newline \bf Lemma 1}. Let Re $a>0$.  Then (a)
$$\sum_{m=1}^\infty {{(-1)^m} \over {m!}}{{B_m(a)} \over m}=-{1 \over a}\left[e^{-a} ~_2F_1(1,
a;a+1;e^{-1})+a(\gamma+\psi(a))\right].  \eqno(1.13)$$
In particular,
$$-\sum_{m=1}^\infty {B_m \over {m!}}{1 \over m}=\sum_{n=1}^\infty {e^{-n} \over n}=1-\ln(e-1).
\eqno(1.14)$$
(b)
$$\sum_{m=1}^\infty {{B_m(a)} \over {m!}}{t^m \over m}=-{1 \over a}\left[e^{at} ~_2F_1(1,
a;a+1;e^t)+a(\gamma+\ln(-t)+\psi(a))\right].  \eqno(1.15)$$  
(c)
$$\sum_{m=1}^\infty {{B_m(a)} \over {m!}}{z^m \over m^2}= -{1 \over a}\int_0^z \left[e^{at} ~_2F_1(1,a;a+1;e^t)+a(\gamma+\ln(-t)+\psi(a))\right]{{dt} \over t}.  \eqno(1.16)$$
(d)
$$\sum_{n=0}^\infty {{\Gamma(0,n+a)} \over {n+a}}=-{1 \over a}\int_0^{1/e} {u^{a-1} \over
{\ln u}} ~_2F_1(1,a;1+a;u)du.  \eqno(1.17)$$


\centerline{\bf Proof of Propositions}

{\it Proposition 1}.  We apply the integration technique Crandall refers to as ``Bernoulli splitting" and use the integral representation for Re $s>0$ (e.g., \cite{nbs}, p. 807)
$$\Gamma(s)\eta(s)=\int_0^\infty {t^{s-1} \over {e^t+1}}dt.  \eqno(2.1)$$
Splitting the integral at $\lambda$, we use the generating function
$${{2e^{xz}} \over {e^z+1}} = \sum_{n=0}^\infty E_n(x) {z^n \over {n!}}, ~~~~|z| < \pi,
\eqno(2.2)$$
for the integration on $[0,\lambda)$.  For the other integration we use a standard
integral representation for the incomplete Gamma function, together with a geometric series
expansion,
$$\int_\lambda^\infty {t^{s-1} \over {e^t+1}}dt=\sum_{m=0}^\infty (-1)^m \int_\lambda^\infty
e^{-(m+1)t}t^{s-1}dt=\sum_{m=0}^\infty {{(-1)^m} \over {(m+1)^s}}\Gamma[s,\lambda(m+1)].
\eqno(2.3)$$ 
The sum in (1.3) converges uniformly away from poles so provides an analytic continuation to 
the left of Re $s>0$.  

For Corollary 1 we recall that for $n \geq 0$ \cite{grad} (p. 941)
$$\Gamma(n+1,x)=n!e^{-x}\sum_{m=0}^n {x^m \over {m!}},  \eqno(2.4)$$
and Li$_2(z)=\sum_{n=1}^\infty z^n/n^2$.


{\it Remarks}.  We have supplied details for Crandall's Algorithm 1 for the $\eta$ function.  
As far as the values $E_n(0)$, their asymptotic form is a case of the following \cite{dlmf} (24.11.6),
\[(-1)^{{\left\lfloor(n+1)/2\right\rfloor}}\frac{\pi^{{n+1}}}{4(n!)}\mathop{E_{{n}}\/}\nolimits\!\left(x\right)\to\begin{cases}\mathop{\sin\/}\nolimits\!\left(\pi x\right),&n\text{ even},\\
\mathop{\cos\/}\nolimits\!\left(\pi x\right),&n\text{ odd},\end{cases}\]
and one has the bounds for $0<x<1/2$ \cite{dlmf} (24.9.5)
\[\frac{4(2n-1)!}{\pi^{{2n}}}\frac{2^{{2n}}-1}{2^{{2n}}-2}>(-1)^{n}\mathop{E_{{2n-1}}\/}\nolimits\!\left(x\right)>0.\]

Similar treatment can be made of 
$$\int_0^\infty {{t^{s-1}e^{-(a-1)t}} \over {e^t+1}}dt=2^{-s}\Gamma(s)\left[\zeta\left(s,{a
\over 2}\right)-\zeta\left(s,{{a+1}\over 2}\right)\right]=\Gamma(s)\sum_{n=0}^\infty {{(-1)^n}
\over {(n+a)^s}},  \eqno(2.5)$$
where Re $a>0$ and Re $s>0$.  This yields, for $|\lambda|<\pi$,
$$2^{-s}\Gamma(s)\left[\zeta\left(s,{a
\over 2}\right)-\zeta\left(s,{{a+1}\over 2}\right)\right]
=\sum_{m=1}^\infty (-1)^m {{\Gamma[s,\lambda(m+a)]} \over {(m+a)^s}}+{1 \over 2}
\sum_{n=0}^\infty {{E_n(1-a)} \over {n!}} {\lambda^{n+s} \over {(n+s)}},  \eqno(2.6)$$
again providing analytic continuation to all of $\mathbb{C}$.  Here the polar part is absent,
and this representation could be used to develop expressions for the differences of
Stieltjes constants $\gamma_k(a/2)-\gamma_k[(a+1)/2]$.

{\it Corollary 2}.  For part (a), use the property $B_m(a+1)=B_m(a)+ma^{m-1}$ and the 
sum (by \cite{grad}, p. 941)
$$\sum_{m=0}^\infty (-1)^{m+1} {a^m \over m!} {\lambda^{m+s} \over {(m+s)}}
=a^{-s}[\Gamma(s,a\lambda)-\Gamma(s)].  \eqno(2.7)$$
For part (b), use the derivative $d\Gamma(\alpha,x)/dx=-x^{\alpha-1}e^{-x}$ and the
functional relations $\alpha\Gamma(\alpha,x)=\Gamma(\alpha+1,x)-x^\alpha e^{-x}$ and 
$\Gamma(s)=\Gamma(s+1)/s$.  For part (c), use $\sum_{r=1}^{q-1}B_m\left({r \over q}\right)
=(q^{1-m}-1)B_m$ and $(-1)^m B_m(1)=B_m$.  The following decomposition, wherein the generating function (2.23) is employed, can then be used to verify the stated property.  
$$q^s\zeta(s)={q^s \over {\Gamma(s)}}\int_0^\infty {y^{s-1} \over {e^y-1}}dy$$
$$={q^s \over {\Gamma(s)}}\left[\int_0^{\lambda/q} {y^{s-1} \over {e^y-1}}dy +
\int_{\lambda/q}^\infty {y^{s-1} \over {e^y-1}}dy\right]$$
$$={1 \over {\Gamma(s)}}\sum_{m=0}^\infty q^{1-m} {B_m \over {m!}}{\lambda^{m+s-1} \over
{(m+s-1)}}+{q^s \over {\Gamma(s)}}\sum_{n=0}^\infty \int_{\lambda/q}^\infty {{e^{-(n+1)qy}
(e^{qy}-1)y^{s-1}} \over {e^y-1}}dy.  \eqno(2.8)$$
For part (d), we first note that $\int_0^1 B_m(a)da=0$ for all $m \geq 1$, giving from (1.1)
$$\Gamma(s)\int_0^1 \zeta(s,a)da=\sum_{n=0}^\infty \int_0^1 {{\Gamma(s,\lambda(n+a)]} \over {(n+a)^s}}da +{\lambda^{s-1} \over {s-1}}.  \eqno(2.9)$$
Integrating by parts,
$$\int_0^1 {{\Gamma(s,\lambda(n+a)]} \over {(n+a)^s}}da=-{1 \over {(s-1)}}\int_0^1 \Gamma(s,
\lambda(n+a)]\left({d \over {da}}{1 \over {(n+a)^{s-1}}}\right)da$$
$$={1 \over {s-1}}\left[-\lambda^{s-1}
e^{-\lambda n}+\lambda^{s-1}e^{-\lambda(n+1)} +n^{1-s}\Gamma(s,\lambda n)-(n+1)^{1-s}\Gamma(
s,\lambda(n+1))\right].  \eqno(2.10)$$
Then the sum in (2.9) telescopes to $-\lambda^{s-1}/(s-1)$ and (1.8) follows.  \qed

{\it Remark}.  More generally than part (c), we have the reciprocity relation (\cite{coffeysums}, Lemma 1) 
$$\sum_{r=1}^q \zeta\left(s,{{pr} \over q}-b\right)=\left({q \over p}\right)^s ~
\sum_{\ell=0}^{p-1} \zeta\left(s,{{\ell q+p} \over p}-{{qb} \over p}\right).  \eqno(2.11)$$
Here $p \geq 1$ and $q \geq 1$ are integers, $b\geq 0$, and min$(p/q,q/p) >b$.  
From (2.11) follows (\cite{coffeysums}, Corollary 3)
$$\sum_{r=1}^q B_m\left({{pr} \over q}-b\right)=\left({q \over p}\right)^{1-m}
\sum_{\ell=0}^{p-1} B_m\left[1+(\ell-b){q \over p}\right].  \eqno(2.12)$$

{\it Proposition 2}.  First integrating by parts we have
$$\Gamma^{(1)}(s,x) \equiv {\partial \over {\partial s}}\Gamma(s,x)=\int_x^\infty t^{s-1}\ln t
~ e^{-t}dt$$
$$=-\int_x^\infty \ln t {\partial \over {\partial t}}\Gamma(s,t)dt
=\int_x^\infty {{\Gamma(s,t)} \over t}dt +\ln x \Gamma(s,x).  \eqno(2.13)$$
We next recall a hypergeometric form of $\Gamma(x,y)$,
$$\Gamma(\alpha,x)=\Gamma(\alpha)-{x^\alpha \over \alpha} ~_1F_1(\alpha,1+\alpha;-x)$$
$$=\Gamma(\alpha)-{x^\alpha \over \alpha}\sum_{j=0}^\infty {\alpha \over {(\alpha+j)}}{{(-x)^j}
\over {j!}},  \eqno(2.14)$$
with $_1F_1$ the confluent hypergeometric function.
This expression may be integrated on a finite interval.  For an improper integral
extending to infinity we need some asymptotic information contained in the following.

{\bf Lemma 2}.  (Asymptotic form of special $_pF_p$ functions)  As $x \to \infty$ (a)
$$_2F_2(s,s;s+1,s+1;-x) \sim e^{-x}{s^2 \over x^2}+x^{-s}s^2\Gamma(s)[\ln x- \psi(s)],
\eqno(2.15)$$
and (b)
$$_3F_3(s,s,s;s+1,s+1,s+1;-x) \sim -e^{-x}{s^3 \over x^3}+x^{-s}{s^3 \over 2}\Gamma(s)
[\ln^2 x-2\ln x \psi(s)+\psi^2(s)+\psi'(s)].  \eqno(2.16)$$
Here $\psi'$ is the trigamma function.

(a) We use a general procedure based upon the Barnes
integral representation of $_pF_q$ (\cite{paris}, Section 2.3).  We have
$$_2F_2\left(s,s;s+1,s+1;-x\right)={1 \over {2\pi i}}\int_{-i\infty}^{i\infty}
\Gamma(-y)\Gamma(y+1)g(y)x^y dy, \eqno(2.17)$$
where the path of integration is a Barnes contour, indented to the left of the origin
but staying to the right of $-s$, and
$$\Gamma(y+1)g(y)={{\Gamma^2(y+s)} \over {\Gamma^2(y+s+1)}}{{\Gamma^2(s+1)} \over {\Gamma^2(s)}}={s^2 \over {(y+s)^2}}.  \eqno(2.18)$$
The contour can be thought of as closed in the right half plane, over a semicircle
of infinite radius.  We then move the contour to the left of $y=-s$, picking up the
residue there.  
We find
$$s^2 \mbox{Res}_{y=-s} ~{{\Gamma(-y)x^y} \over {(y+s)^2}}={s^2 \over x^s}\Gamma(s)[\ln x-\psi(s)].  \eqno(2.19)$$
This gives the algebraic part of the asymptotic form of the $_2F_2$ function, that is the leading portion.  The higher order terms come from the exponential expansion of these functions, and they are infinite in number.  The latter expansion may be determined according to (\cite{paris}, Section 2.3, p. 57).

(b) proceeds similarly, where now
$$s^3 \mbox{Res}_{y=-s} ~{{\Gamma(-y)x^y} \over {(y+s)^3}}={s^3 \over {2x^s}}\Gamma(s)
[\ln^2 x-2\ln x \psi(s)+\psi^2(s)+\psi'(s)].  \eqno(2.20)$$

We may integrate in (2.11) from $x$ to $b$ using (2.12),
$$\int_x^b {{\Gamma(s,t)} \over t}dt={x^s \over s^2} ~_2F_2(s,s;s+1,s+1;-x)-{b^s \over s^2} ~_2F_2(s,s;s+1,s+1;-b)+\Gamma(s)(\ln b-\ln x).  \eqno(2.21)$$
Then with the Lemma in hand, taking $b \to \infty$ completes the Proposition.  \qed  

{\it Proposition 3}.  We take $\lambda=1$ in (1.1) and, making use of Proposition 2 and
related considerations, expand the function
$$\Gamma(s)\zeta(s,a)-{1 \over {s-1}}=\sum_{n=0}^\infty {{\Gamma(s,n+a)} \over {(n+a)^s}}+ \sum_{m=1}^\infty {{(-1)^m} \over {m!}} {{B_m(a)} \over {(m+s-1)}} \eqno(2.22)$$
about $s=1$, and apply the defining series (1.2).  

{\it Corollary 3}.  Maintaining the free parameter $\lambda$, this follows from
$-\int_1^x[\gamma+\psi(a)]da=\gamma(1-x)-\ln \Gamma(x)$, the corresponding expression
similar to (1.10), 
$$-\ln \lambda-\gamma-\psi(a)=e^{-\lambda a}\Phi\left({1 \over e^\lambda},1,a\right)+\sum_{m=1}^\infty {{(-1)^m} \over {m!}}{{B_m(a)} \over m}\lambda^m, 
\eqno(2.23)$$
and the property $\int B_m(a)da=B_{m+1}(a)/(m+1)$.

{\it Lemma 1}.  For (a)-(c), repeatedly integrate the generating function
$$\sum_{n=1}^\infty {{B_n(a)} \over {n!}}z^n = {{ze^{az}} \over {e^z-1}}-1, ~~~~|z| <2\pi.
\eqno(2.24)$$
For (d), write
$$\sum_{n=0}^\infty {{\Gamma(0,n+a)} \over {n+a}}=\sum_{n=0}^\infty {1 \over {n+a}}
\int_{n+a}^\infty e^{-t} {{dt} \over t}=\sum_{n=0}^\infty {1 \over {n+a}}\int_1^\infty
e^{-(n+a)v} {{dv} \over v}.  \eqno(2.25)$$
Then interchange summation and integration and perform another change of variable.
\qed

{\it Remark}.  As $\Gamma(0,x)=-\mbox{Ei}(-x)$, where Ei is the exponential integral
having many integral representations, there are a multitude of ways of obtaining (1.17).

\medskip
\centerline{\bf Discussion:  Derivatives of the incomplete Gamma function}
\medskip

Derivatives of the incomplete Gamma function are important for the computational methods of
\cite{crandall2012}, and they provide a starting point for many useful integrals.
Geddes et al.\ \cite{geddes} investigated these derivatives, introducing a function $T(m,a,z)$, with $\Gamma(a,z)=zT(2,a,z)$, $m \geq 1$ an integer and initially $|z|<1$, such that
$$\Gamma^{(1)}(a,x)=xT(3,a,x)+\ln x ~\Gamma(a,x), \eqno(3.1a)$$
$$\Gamma^{(2)}(a,x)=\ln^2 x ~\Gamma(a,x)+2x[\ln x ~T(3,a,x)+T(4,a,x)], \eqno(3.1b)$$
and more generally, with $P_j^i=i!/(i-j)!$, 
$${{d^m \Gamma(a,x)} \over {da^m}}\equiv \Gamma^{(m)}(a,x)=\ln^m x ~\Gamma(a,x)
+mx \sum_{i=0}^{m-1} P_i^{m-1} \ln^{m-i-1}x ~T(3+i,a,x).  \eqno(3.2)$$
The function $T$ satisfies the derivative relations
$${{dT(m,a,z)} \over {da}}=\ln z T(m,a,z)+(m-1)T(m+1,a,z), \eqno(3.3a)$$
and
$${{dT(m,a,z)} \over {dz}}=-{1 \over z}[T(m-1,a,z)+T(m,a,z)].  \eqno(3.3b)$$
One form of $T$ is \cite{geddes} (37) 
$$T(m,a,z)=-\mbox{Res}_{s=-1}\left[\left(- {1 \over {s+1}}\right)^{m-1} \Gamma(a-1-s)z^s\right]
+\sum_{i=0}^\infty {{(-1)^i z^{a+i-1}} \over {i!(-a-i)^{m-1}}},  \eqno(3.4)$$
where $a$ is neither zero or a negative integer.  This can be obtained by expressing $T$ in
terms of the Meijer G-function and then using a contour integral for the latter function.
Using \cite{geddes} (38) and the relation $(a)_i/(a+1)_i=a/(a+i)$, as we have previously done, we may write
$$T(m,a,z)={{(-1)^{m-1}} \over {(m-2)!}} \left({d \over {dt}}\right)^{m-2} \left[\Gamma(a-t)
z^{t-1}\right]_{t=0}$$
$$+(-1)^{m-1} {z^{a-1} \over {a^{m-1}}} ~_{m-1}F_{m-1}(a,a,\ldots,a;a+1,a+1,\dots,a+1;-z).  \eqno(3.5)$$
Proposition 2 and like considerations are in accord with this result.

\pagebreak
\centerline{\bf Certain zeta function and other integrals}
\medskip

Elsewhere we have considered second moment integrals of the Riemann zeta and other functions
\cite{coffeymoments}.  The following concerns the integrals
$$I(c) \equiv \int_{-\infty}^{\infty} {{\zeta(c+it)} \over {c+it}}dt
=-i\int_{\mbox{\tiny{Re}}s=c}{{\zeta(s)} \over s}ds,  \eqno(4.1)$$
$$I_a(c) \equiv \int_{-\infty}^{\infty} {{\eta(c+it)} \over {c+it}}dt
=-i\int_{\mbox{\tiny{Re}}s=c}{{\eta(s)} \over s}ds,  \eqno(4.2)$$
and
$$I_L(c) \equiv \int_{-\infty}^{\infty} {{\mbox{Li}_{c+it}(x)} \over {c+it}}dt
=-i\int_{\mbox{\tiny{Re}}s=c}{{\mbox{Li}_s(x)} \over s}ds.  \eqno(4.3)$$

\noindent
{\bf Proposition 4}.  We have $I(c)=-\pi$ for $0<c<1$, and
$I(c)=\pi$ for $c>1$,  
$I_a(c=0)=2\pi$ and $I_a(c)=\pi$ for $c>0$, and $I_L(c)=\pi x$ for $|x|<1$, $x \in \mathbb{R}$ and $c>0$ and $I_L(c)=-\pi x(1+x)/(1-x)$ for $c<0$.




{\it Proof}.  These results may be developed with the aid of Perron's formula \cite{apostol} 
(p. 245), one form of which is \cite{crandallpom} (p. 147)
$${1 \over {2\pi i}}\int_{\mbox{\tiny{Re}}s=\sigma} \left({x \over n}\right)^s {{ds} \over s}
=\theta(x-n), \eqno(4.4)$$
where $\theta(x)$ is the step function taking values $1$, $1/2$, and $0$ as $x>0$, $x=1$,
and $x<0$, respectively.  The line of integration may be moved by including the poles of
the integrand.  This may be seen by integrating over a rectangular contour with corners at
$c \pm iT$ and $b \pm iT$.  For the Hurwitz zeta function one has the growth estimates
\cite{ww} (p. 276), with $s=\sigma+it$ and $1/2>\delta>0$, 
$$\zeta(s,a)=O(1) ~\mbox{for} ~\sigma>1+\delta, ~~~~~~$$
$$          =O(|t|^{1/2-\sigma}) ~\mbox{for} ~\sigma \leq \delta, $$     
$$          =O(|t|^{1/2}) ~\mbox{for} ~\delta \leq \sigma \leq 1-\delta, $$ 
$$          =O(|t|^{1-\sigma}\ln |t|) ~\mbox{for} ~1-\delta \leq \sigma \leq 1+\delta, $$    
$$          =O(|t|^{1/2}\ln |t|) ~\mbox{for} ~-\delta \leq \sigma \leq \delta.  \eqno(4.5)$$     
Further inequalities for $\zeta_a(1-s)=\sum_{n=1}^\infty e^{2n\pi i a}/n^{1-s}$ are also
given in \cite{ww}.  Cases of these may be used to show that the `top' and `bottom'
contributions of the rectangular contour integration vanish as $T \to \infty$.  Then the
Cauchy residue theorem is applied.  Example estimations along horizontal line segments 
include the following, wherein we note the independence of the parameter $a$, and $c_1$ and
$c_2$ are positive constants.
$$\left|\int_{-\delta}^\delta {{\zeta(\sigma \pm iT)} \over s}d\sigma \right|  
\leq \int_{-\delta}^\delta {{|\zeta(\sigma \pm iT)|} \over s}d\sigma $$
$$\leq {1 \over T}\int_{-\delta}^\delta |\zeta(\sigma \pm i T)|d\sigma
\leq {c_1 \over T}\int_{-\delta}^\delta |T|^{1/2} \ln |T|d\sigma =2\delta {c_1 \over T^{1/2}}
\ln |T|,$$
and
$$\left|\int_{1-\delta}^{1+\delta} {{\zeta(\sigma \pm iT)} \over s}d\sigma \right|  
\leq \int_{1-\delta}^{1+\delta} {{|\zeta(\sigma \pm iT)|} \over s}d\sigma $$
$$\leq {1 \over T}\int_{1-\delta}^{1+\delta} |\zeta(\sigma \pm i T)|d\sigma
\leq {c_2 \over T}\int_{1-\delta}^{1+\delta} |T|^{1-\sigma} \ln |T|d\sigma$$
$$\leq {{c_2 \ln T} \over T}\int_{1-\delta}^{1+\delta}|T|^{1-\sigma} d\sigma
={c_2 \over T}(T^\delta -T^{-\delta}).$$
The interval $[c,b]$ can then be appropriately decomposed with such contributions.  

The relevant residues are the following.  For $\eta(s)/s$, $1/2$ at $s=0$.  For $\zeta(s)/s$,
$-1/2$ at $s=0$, being a case of $\zeta(0,a)=1/2-a$, and $1$ at $s=1$.  For Li$_s(x)/s$, $x/(1-x)$ at $s=0$.

So for instance, for (4.1) with $c>1$, the sum on $n$ in (4.4) may be performed.
For $c<1$ the contribution from the pole at $s=1$ is subtracted.

Similarly for (4.3), when $c<0$, the stated formula obtains from $2\pi\left({x \over 2}
-{x \over {1-x}}\right)$.


{\it Remark}. It appears the integrals $\int_{-\infty}^{\infty} {{\zeta(c+it)} \over {(c+it)^p}}dt$ for $p \geq 1$ and $0<c<1$ take only the values $-\pi$ or $-2\pi$. 


\medskip
\centerline{\bf Acknowledgement}
I thank R. Crandall for reading the manuscript, and in particular, for assisting with
Section 4.

\pagebreak

\end{document}